\documentclass[preprint,prd,tightenlines,superscriptaddress,showkeys]{revtex4}

\usepackage{graphicx} 
\usepackage{dcolumn}  
\usepackage{colordvi}
\usepackage{color}
\usepackage{epstopdf}
\usepackage{subfig}
\usepackage{amssymb}
\usepackage{url}
\graphicspath{{ps}}
\usepackage{hyperref}
\usepackage{tabularx}

\input belle2sym.tex

\def \Y       {\ensuremath{\Upsilon(4S)}\xspace}

\begin{document}


\def\belletwo {{Belle II}\xspace}
\def\itbelletwo {{\it {Belle II}}\xspace}
\def\phaseiii {{Phase III}\xspace}
\def\itphaseiii {{\it {Phase III}}\xspace}

\newcommand\logten{\ensuremath{\log_{10}\;}}

\vspace*{-3\baselineskip}
\resizebox{!}{3cm}{\includegraphics{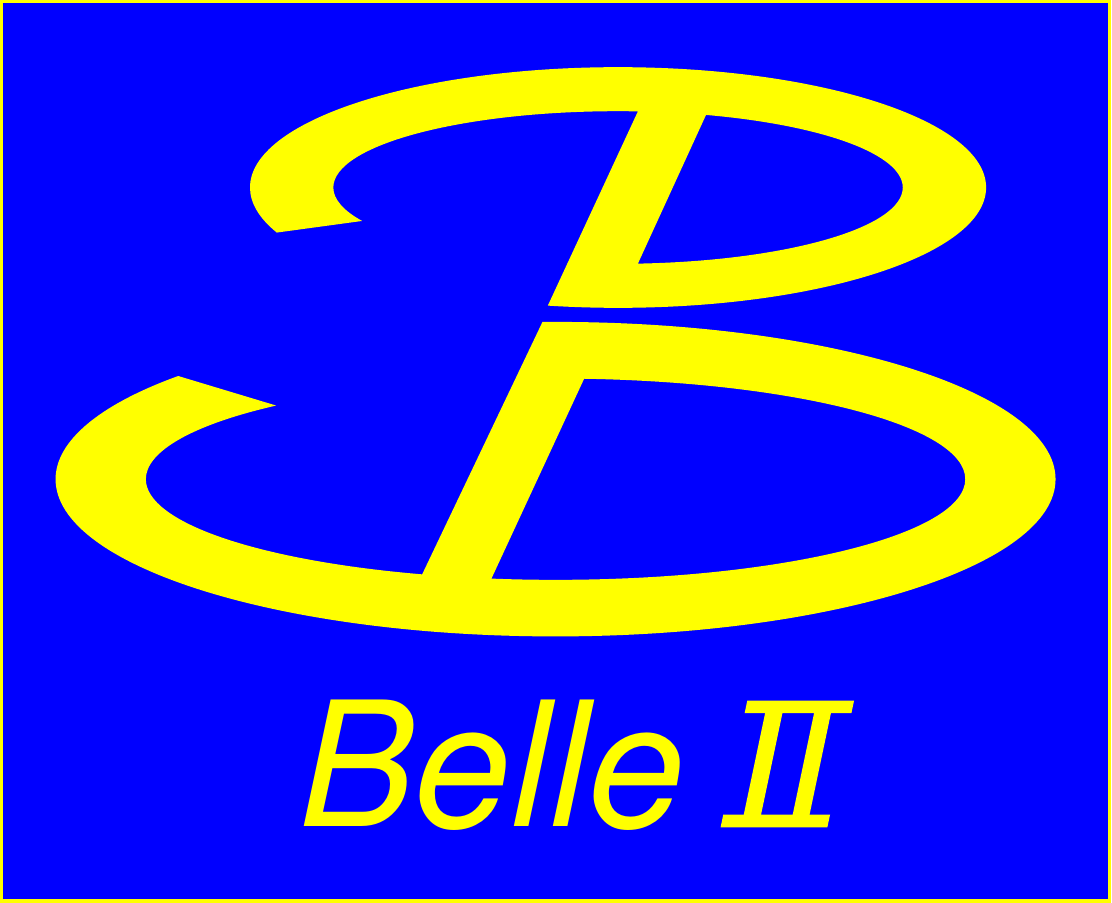}}

\vspace*{-5\baselineskip}
\begin{flushright}
BELLE2-CONF-PROC-2020-018\\
\today
\end{flushright}

\title { \quad\\[0.5cm] Diversity + Inclusion at Belle II:\\ Where We Are, What We've Done and Where We Want To Be}
\altaffiliation{\textit{Presented at 40th International Conference on High Energy physics - ICHEP2020; July 28- August 6, 2020; Prague, Czech Republic (virtual meeting)}}

\author{S.A.~De La Motte}
\thanks{Speaker}
\email{shanette.delamotte@adelaide.edu.au}
\affiliation{Department of Physics, The University of Adelaide,\\ Adelaide South Australia 5005 Australia}
\author{H.M.~Wakeling}
\affiliation{Department of Physics, McGill University,\\
3600 rue University, Montr\'eal, Canada}

\author{M.~Barrett}
\affiliation{Institute of Particle and Nuclear Studies\\
  High Energy Accelerator Research Organization (KEK),\\
  1--1 Oho, Tsukuba, Ibaraki, 305--0801, Japan}

\author{K.~Kinoshita}
\affiliation{University of Cincinnati\\  Cincinnati, Ohio, 45221, U.S.A}

\collaboration{The Belle II Collaboration}
\noaffiliation

\begin{abstract}
The Belle II Collaboration comprises over 1000 international high energy
physicists, who investigate the properties of $b$-quarks and other
particles at the luminosity frontier.
In order to achieve our aim of a successful physics program, it is essential that we emphasise contributions from a diverse community. Belle II has thus far focused on diversity in gender and sexuality, among other efforts within our collaboration. These efforts are led by our two Diversity Officers, elected to the newly created positions in 2018. Their role has been to promote
an inclusive atmosphere, raising awareness of diversity and being a
safe first point of call for issues of discrimination and harassment.

These proceedings accompany the short talk delivered during ICHEP 2020 \cite{slides}, marking the first conference the Belle II Collaboration has presented in the diversity and inclusion stream. It details the efforts described above, as well as examining the evolving gender demographics of our community, since membership began in 2011.\\
\end{abstract}

\maketitle

{\renewcommand{\thefootnote}{\fnsymbol{footnote}}}
\setcounter{footnote}{0}

\section{What is meant by diversity and inclusion?}
In order to motivate the importance of diversity and inclusion it is necessary to first carefully define what these words mean. \textit{Diversity} describes the large range of differences from person to person within society. Examples of this can be differences in gender, race, sexual orientation, ability or socio-economic status. Through personal experience or statistics taken within the field, high energy physicists should be well aware that the diversity of the wider community is not necessarily be replicated within their collaborations.
Thus, \textit{inclusion} describes actions that can be taken to support those from under-represented groups, to provide these collaborators a fair opportunity to do good physics.
\section{Who we are}
The Belle II Collaboration primarily seeks to make measurements of $B$-physics at both the luminosity and precision frontier. Even though data is taken using an electron-positron collider located in Japan, the Collaboration currently has 1041 members across 26 countries. Belle II is also unique amongst other larger high energy physics collaborations due to the large number of participants from Asia -- 35\% of members belong to institutes in Asian countries.
While diverse in the regions of its institutes, this same diversity is unfortunately not seen in the gender of our collaborators. In 2011 membership records, approximately 12\% of collaborators identified as female. As of 2019 this proportion is 15\%, a small increase given that the total number of collaborators has doubled in this time.
\begin{figure}
    \centering
    \includegraphics[scale=0.5]{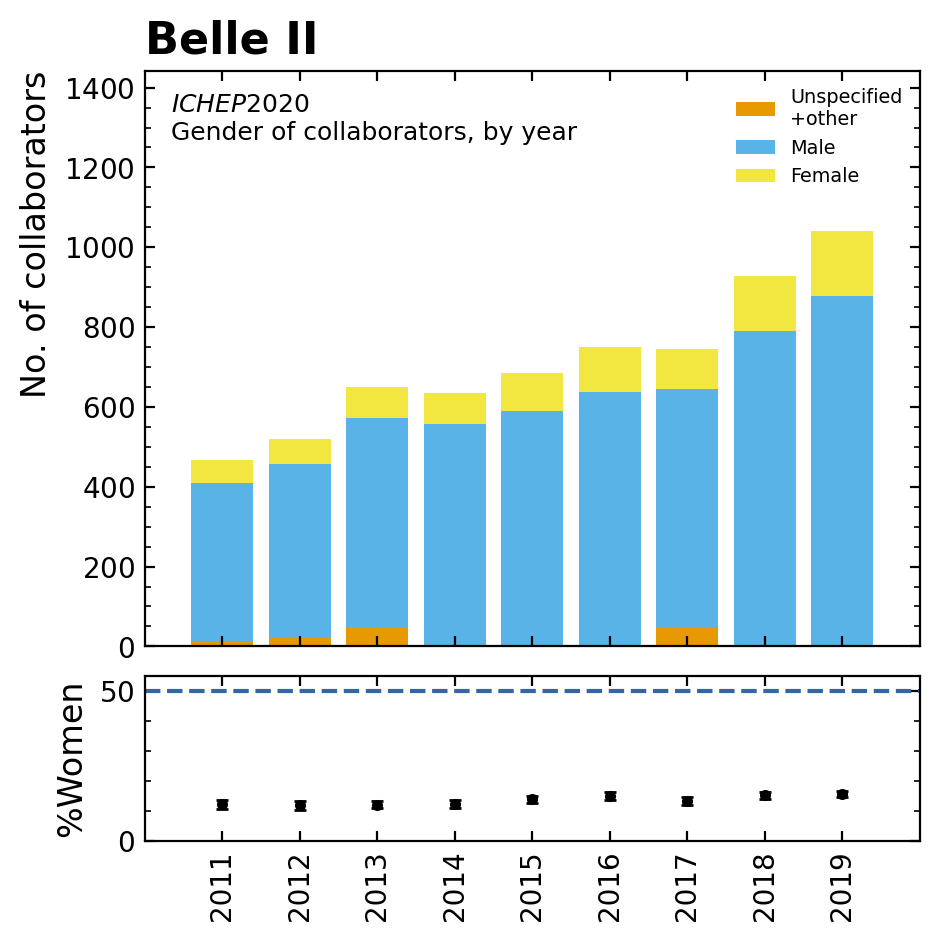}
    \caption{Gender demographics of Belle II collaborators. Data taken from Belle II membership declaration, where `other' refers to non-binary gender identity and `unspecified' refers to undeclared gender information.}
    \label{fig:my_label}
\end{figure}
\section{What we have done}
In order to make the Collaboration aware of this disparity in gender and other efforts in inclusion, the role of Diversity Officer was created in 2018. The role of the two Belle II Diversity Officers is to offer guidance on topics of discrimination and harassment within the collaboration, as well as to hold open forums for members to voice concerns during Belle II general meetings. Matt Barrett and Kay Kinoshita are the current members fulfilling this position. One of their first initiatives was to gather poll data on what aspects of inclusion were important to collaborators. 
26\% of responses agreed with the statement that they did not pursue leadership due to the impact the workload could have on their family life. 
One response to this was to compile information on how visitors can access childcare during on-campus general meeting periods, to encourage participation from international collaborators with families. 
In order to further promote inclusion in the collaboration, the Diversity Officers have provided resources in the internal wiki to make official Belle II plots accessible for those with colour-blindness, which were used to make the plot in figure \ref{fig:my_label}. They have also provided advisories on the use of language that may exclude under-represented groups, such as avoiding the gendered connotations in terms such as `manpower'. Commitment to normalising inclusive language has also taken place within our computing group, with respect to words with racial overtones. Following a similar decision by GitHub \cite{github}, we are removing references to `master' and `slave' in Belle II software.

The Belle II social media accounts have been used to publicly support under-represented groups in Physics, through the International Day of Women and Girls in Science and LGBTSTEM day. Belle II remains the only non-US and non-European organisation to demonstrate support for LGBTSTEM day and aims to encourage other Japanese STEM organisations to do the same.

At the location of the Belle II experiment itself, we've sought to make the working areas more inclusive. This includes availability of bathrooms for women and for those who identify as gender neutral in high contact areas such as the dormitories and control rooms.

\section{Where we want to be}
The inclusion efforts at Belle II are still a work in progress -- much more is needed to actively encourage and maintain diversity in collaborators. In addition to what has been discussed above, future efforts are in discussion on how Belle II can support differences in race and mental health. It is the hope of the Belle II Collaboration that future presentations at conferences can demonstrate improvements not only in analysis datasets but also in initiatives to improve our demographics. Support for under-represented groups can result in more diverse approaches to analysis, and is thus a valuable way to do good, unbiased physics.

\bibliography{belle2}
\bibliographystyle{belle2-note}

\end{document}